\documentclass{pasj01}
\begin{document}
\SetRunningHead{K. Sadakane et al.}{Emission lines in early B-type stars}
\Received{}
\Accepted{}

\title{Weak Metallic Emission Lines in Early B-Type Stars} 

\author{Kozo \textsc{Sadakane},\altaffilmark{1}
         and
        Masayoshi \textsc{Nishimura}\altaffilmark{2}}

\altaffiltext{1}{Astronomical Institute, Osaka Kyoiku University, Asahigaoka,  Kashiwara-shi, 
            Osaka   582-8582}
\email{sadakane@cc.osaka-kyoiku.ac.jp}

\altaffiltext{2}{2-6, Nishiyama-Maruo, Yawata-shi, Kyoto  614-8353 }
\KeyWords{Stars:  B-type--- Stars: atmosphere --- Stars: emission line--- Stars: spectroscopy}

\maketitle

\begin{abstract}

Previously unrecognized weak emission lines originating from high excitation   
states of Si~{\sc ii} (12.84 eV) and Al~{\sc ii}  (13.08 eV) are detected 
in the red region spectra of slowly rotating early B-type stars. We
surveyed high resolution spectra of 35 B-type stars covering spectral sub-types
between B1 and B7 near the main sequence
and found the emission line of Si~{\sc ii}  at 6239.6 \AA~ in all 13 stars
having spectral sub-types B2 and B2.5. There are 17 stars belonging to sub-type  B3 and
seven stars among them are found to show the  emission line of Si~{\sc ii}. 
The emission line of Al~{\sc ii} at 6243.4 \AA~ is detected 
in a narrower temperature range ({\it $T_{\rm eff}$}  between 19000K and 23000 K) in nine stars. 
Both of these emission lines are not detected in cooler 
({\it $T_{\rm eff}$}  $ < $ 16000 K) stars in our sample.
The emission line of Si~{\sc ii}  at 6239.6 \AA~ shows a single-peaked and 
symmetric profile and the line center has no shift in wavelength with respect to 
those of low excitation absorption lines of Si II. Measured half width of the 
emission line is the same as those of rotationally broadened low excitation 
absorption lines of Si~{\sc ii}.
These observations imply that the emitting gas is not circumstellar origin,
but is located at the outermost layer of the atmosphere, covering the whole 
stellar surface and co-rotates with the star.

\end{abstract}

\section{Introduction}

The phenomenon of sharp and weak emission lines (WELs) in optical
spectra of middle to late B-type stars has been reported mainly in
chemically peculiar (CP) stars. Detections of these weak features
have been made possible by achieving both high spectral resolution 
and high signal-to-noise (SN) ratio observations.
\citet{sigut2000} reported detections of emission lines of Mn~{\sc ii}, 
P~{\sc ii}, and Hg~{\sc ii}  in the red spectral region  of the helium-weak 
star 3 Cen A (B5 III-IVp). They also found very weak emission lines
of Mn~{\sc ii}  in a mild HgMn star 46 Aql (B9 III).
Sigut (2001a, b)  carried out detailed analyses of emission lines of Mn~{\sc ii}  
observed in 3 Cen A and 46 Aql and concluded that observations can be 
naturally explained by interlocked non-local thermodynamic equilibrium 
(NLTE)  effect combined with the vertical stratification of the manganese 
abundance,  with manganese concentrated high in the photosphere.


\citet{wahlgren2000} reported detections of weak emission lines 
originating from high excitation states of Ti~{\sc ii}, Cr~{\sc ii}  and Mn~{\sc ii} in 
sharp lined late
B-type stars including HgMn stars and suggested that the these
emission lines arise from a selective excitation process involving H Ly$\alpha$
 photon energies.
\citet{sadakane2001} noted that all of the Ti~{\sc ii} lines in  46 Aql  with high excitation potential 
($\chi$  $ > $ 8.05 eV) and large transition probabilities (log $\it gf$
 $ > $  0.1) are observed  in emission near 6000 \AA.
\citet{wahlgren2004} published an extensive list of emission lines of 
3 Cen A. Their list includes emission line spectra of Si~{\sc ii}, P~{\sc ii}, Ca~{\sc ii},
Mn~{\sc ii}, Fe~{\sc ii}, Ni~{\sc ii}, Cu~{\sc ii}, and Hg~{\sc ii}. Abundances of 11 elements 
(from C to Hg) have been determined using a synthetic spectrum fitting 
technique.       

Numerous weak emission lines of Cr~{\sc ii} and Ti~{\sc ii} have been reported 
in a cool HgMn star  HD 175640 (B9 V) by  \citet{castelli2004}.
They noted that emission lines are selectively found  for
high excitation lines having large transition probabilities 
(log $\it gf$  $ > $ --1.0) and that these emission lines are found 
in the red part ($\lambda$  $ > $  5850 \AA) of the spectrum.
\citet{castelli2007} observed emission lines of Cr~{\sc ii}, Mn~{\sc ii}  and Fe~{\sc ii} 
and found isotopic anomalies for Ca and Hg in the Bp star HR 6000. 
They noted that there are disagreements between
observed and calculated line strengths for some Fe~{\sc ii}  lines and
 suggest these discrepancies are due either to incorrect
log $\it gf$ values or to the emission component filling the 
absorptions (filling-in effects).

\citet{hubrig2007} observed high resolution spectra of the sharp-lined magnetic 
helium-variable star a Cen (HD 125823) over the rotation period of 8.82 d 
and found  variable high excitation S~{\sc ii}, Mn~{\sc ii} and Fe~{\sc ii} emission lines.
They found a correlation between the probable location of surface spots of Mn and
Fe and the strength of the emission lines.
It is interesting to notice that they note that an emission line 
detected in this star at 6239.80 \AA~ as an unidentified line (their table 1). 
Recently, \citet{alexeeva2016} constructed a comprehensive model atom for C~{\sc i} 
and C~{\sc ii} and computed the NLTE  line 
formation for C~{\sc i} and C~{\sc ii}. They analysed the lines of C~{\sc i} and 
C~{\sc ii} in seven  B to early A-type stars, including $\iota$ Her, Sirius, and Vega, 
and found that the C~{\sc i} emission lines were detected in the
four hottest stars, and these lines were well reproduced by their NLTE calculations. 

Summarizing these published results, we notice that stars
showing WELs are mainly middle B to early A-type stars including CP stars. 
Almost all WELs are  arising from  high excitation states of singly ionized ions.
At the same time, WELs are found in the red and near-IR spectral regions 
preferentially and  no WELs have been reported in the blue or in the
near UV regions so far. 

\citet{nieva2011} analysed high resolution spectra of 13 early B-type 
main-sequence stars of spectral classes B0 V to B2 V and low projected rotational velocities
($\it v$ sin $\it i$ $ < $ 60 km s${}^{-1}$) in the Ori OB1 association. They published a
graphical presentation of the spectrum of HD 35299 (B1.5 V) in the appendix.
\citet{nieva2012} carried out a comprehensive spectral analyses of 27 B-type stars
and published graphical presentations of spectra of four B-type stars including a 
B2 IV star $\gamma$ Peg.
Examining graphical data of HD 35299 \citep{nieva2011} and
 $\gamma$ Peg \citep{nieva2012}, we noticed an
 emission  like spike near 6239.7 \AA~ in both stars. Curiously, however, no
identification has been given to this feature on graphs of both stars.

Interested in the nature of this unidentified feature, we examined  high 
resolution spectral data of $\gamma$ Peg obtained with the HIgh-Dispersion Echelle 
Spectrograph (HIDES, $\it R$ $\sim$ 70000) at the coud\'e focus of the 188-cm reflector of
the Okayama Astrophysical Observatory (OAO) and data 
downloaded from the Elodie archive ($\it R$ $\sim$ 42000, \cite{moultaka2004}). We  
find that the emission like
feature is definitely present on both data and conclude that the feature is
not an observational artifact. We surveyed for candidate lines by simulating the spectrum
and found two highly excited (12.84 eV) lines of Si~{\sc ii}   at 6239.61 and 6239.66 \AA~ 
are expected to appear as weak absorption lines in early B-type stars. 
Measured wavelength of the emission line just coincides with those of the two Si~{\sc ii} 
lines. Furthermore, we found two additional weak emission features at 6231.8 \AA~ and at 
6243.4 \AA~ in $\gamma$ Peg. Observed wavelengths of these two features 
coincide with those of two highly excited (13.08 eV) lines of Al~{\sc ii} at 
6231.75 \AA~ and at 6243.37 \AA.    

Because we can find no alternative identification for these features,  
we  conclude that WELs of Al~{\sc ii} and Si~{\sc ii} are present 
in the optical region spectrum of $\gamma$ Peg. This finding is the 
first case of WELs found not only in  $\gamma$ Peg but also  in hot 
({\it $T_{\rm eff}$}  $ > $ 20000 K)  B-type stars.

 We collected as many  high resolution spectral data of early
B-type stars as possible from various data archives, then tried a systematic 
survey for the Al~{\sc ii}  and Si~{\sc ii}  emission lines near 6240 \AA~ in these stars
to find that the Si~{\sc ii}  emission line appears frequently in B2 - B3 sharp-lined 
main sequence stars while the Al II emission line appears in a narrower spectral
range of B2-type stars.   
Details of our survey for WELs in early B-type stars and results are described in the 
following sections. 

\section{Observational data}

We collected optical spectral data of 35 B-type stars 
of low rotational velocities corresponding to
spectral types from B1 to B7 and luminosity classes from III to V from 
various data archives. 
Spectral data of 26 targets were observed using the HIDES spectrograph at OAO. 
Data of 11 targets were obtained
with the Elodie spectrograph of the 193-cm telescope at the Observatoire 
de Haute Provence. Data were obtained by the 
UVES spectrograph of the ESO VLT (three targets)  and by the  High Dispersion
Spectrograph (HDS)  of the Subaru telescope (three targets). Raw data 
obtained by the HIDES and HDS spectrographs were downloaded from
the SMOKA database, which is operated by the Astronomy Data Center, 
National Astronomical Observatory of Japan.
 Processed data of the Elodie and UVES spectrographs
were obtained from the Elodie archive
and the UVES-POP archive \citep{bagnulo2003}, respectively. 

The reduction of two-dimensional echelle spectral data obtained 
by the HIDES and HDS spectrographs  (bias subtraction,
flat-fielding, scattered-light subtraction, extraction of spectral data,
and wavelength calibration) was performed in a standard manner using 
the IRAF software package.
 The wavelength calibration was done using the Th-Ar
comparison spectra obtained during the observations.  The observed wavelengths
of all stars observed by all four spectrographs 
have been converted into the laboratory scale using measured wavelengths of 
He~{\sc i}  and Si~{\sc ii}  absorption lines.  Finally, the continuum levels of 
all spectral data have been  normalized to unity by a polynomial fitting technique.

 Relevant data of our target objects are summarized in table 1. 
Spectral types are taken from the Bright Star Catalog 5-th edition \citep{hoffleit1991},
except for three stars (HD 89587, HD 133518 and HD 181858). Spectral types of  two 
southern stars (HD 89587 and HD 133518) are taken from \citet{houk1978} and data
of HD 181858 is taken from \citet{houk1999}. Data of rotational velocities
($\it v$ sin $\it i$) are taken from \citet{abt2002}. Data of $\it v$ sin $\it i$ for
 five stars (HD 3360, HD 29248,
HD 35039, HD35468, and  HD36591) have been replaced with new data taken from
\citet{simon2014} and that of  HD 133518 is taken from  \citet{alecian2014}.
We could find no published data of  $\it v$ sin $\it i$  for HD 89587.

Effective temperatures ({\it $T_{\rm eff}$}) and surface gravities (log $\it g$) are
taken from various sources. We adopt data given in \citet{takeda2010} and 
\citet{nieva2013}. When no data can be found in these two papers, we 
use {\it $T_{\rm eff}$}  and log $\it g$ values given in \citet{lefever2010}  (HD 35468 and HD
214993), \citet{aerts2014} (HD 163472),  \citet{morel2008} (HD 170580), and
\citet{alecian2014} (HD 133518).
We can find neither published data of   {\it $T_{\rm eff}$}  and log $\it g$  nor 
$\it uvby$ and $\beta$ photometric data for the southern star HD 89587. We 
estimated its effective temperature by comparing the line intensity ratio 
of the  S~{\sc ii} line at 5664.8 \AA~  to the  N~{\sc ii} line at 
5666.6 \AA~ with those measured in 15 B2 and B3 type stars in table 1. The ratio  
is found to be sensitive to a  change in temperature for stars 
belonging to spectral types B2 and B3.
Measured equivalent widths (in m\AA)  and central relative intensities (C. I.)
of  the Si~{\sc ii} line  at 6239.6 \AA~ and 
the Al~{\sc ii} line  at 6243.4 \AA~ are given in columns from  12  to 15.
Measurements of equivalent widths are carried out using the direct integration method.

Figures 1 and 2 show spectral data near 6240 \AA~ for  five representative
examples of hot B-type stars and five cooler stars, respectively. We plot observed
 and simulated spectra for each star. Simulated spectra were computed using
the tabulated atmospheric parameters and the rotational velocity for each star, interpolating  
 the ATLAS9 model atmospheres \citep{kurucz1993}.
The solar abundances,  the LTE line formation and the microturbulent velocity, $\xi$${}_{\rm t}$ 
= 4 km s${}^{-1}$,  have been assumed in these simulations.
   We use log $\it gf$ values  of the 
 Si~{\sc ii} lines and the Al~{\sc ii} lines taken from  the NIST Atomic 
Database \citep{kramida2015}, while  log $\it gf$ values  of other lines are taken from
\citet{kurucz1995}. 

We surveyed for other emission features of high excitation Al~{\sc ii} and Si~{\sc ii} ions 
on a high SN ratio ($\sim$ 900) spectrum of $\gamma$ Peg and found two 
 and six  additional emission features of these two ions, respectively, as listed in table 2. 
Figure 3 displays two small sections of the spectra of two sharp lined stars 
$\gamma$ Peg  and $\iota$ Her near the
 Si~{\sc ii} line at 5688.81 \AA~ (upper panel) and the  Al~{\sc ii} line at 
5593.30 \AA~(lower panel). We can see that the Si~{\sc ii} line is seen
in emission in both stars. The Al~{\sc ii} line appears as an emission
feature in $\gamma$ Peg, while the line is seen as a weak absorption feature 
in the cooler star $\iota$ Her.  

In figure 4, we compare observed profiles of the  Si~{\sc ii} emission line 
 at 6239.6 \AA~ and those of  a low excitation (8.12 eV) absorption line of 
 Si~{\sc ii} at 6371.37 \AA~  of three
B2 type stars ($\gamma$ Peg,   $\zeta$ Cas, and $\gamma$ Ori)  
on the velocity scale.

\section{Discussion}

We have presented observations of weak emission lines (WELs) of high
excitation Si~{\sc ii} and Al~{\sc ii} ions in the red region spectra of
early B-type stars.  WELs in hot B-type stars and those of 
Al~{\sc ii} ions  have not been reported  in previous publications.
We find that the emission line of Si~{\sc ii} at 6239.6 \AA~ is observed
in all 13 stars in our sample contained in the  {\it $T_{\rm eff}$}  range between
23000 K and 17500 K. The line is observed less frequently in middle B-type  
stars (between 17500 K and 16500 K) and the line is not observed in emission
among cooler ({\it $T_{\rm eff}$} $ < $ 16000 K) stars.  The observed high frequency of 
detection of the  Si~{\sc ii} line at 6239.6 \AA~ in emission among early B-type
stars strongly suggests that the occurrence of this emission line is not a rare 
case but a common phenomenon among these stars.
The line of Al~{\sc ii} at 6243.4 \AA~ is observed in emission only in a
narrow temperature range  (between 23000 K and 19000 K) less frequently than
the Si~{\sc ii} line. The line is observed not in emission but as a weak 
absorption feature in stars with  {\it $T_{\rm eff}$}  $ < $ 17500 K.
Data of equivalent widths of emission lines (table 1, columns 12 and 14) show that
maxima of both emission lines are observed between the range in   {\it $T_{\rm eff}$}  
between 22000 K and 19500 K.

We have data of multi-epoch observations for two objects $\gamma$ Peg  and
$\iota$ Her,  both of which show a clear emission line of  Si~{\sc ii} at 
6239.6 \AA. Data of six epochs (from 1998 November to 2006 October) and seven
epochs (from 1994 April to 2013 June) are available for $\gamma$ Peg   and $\iota$ Her,
respectively. These data are obtained by the Elodie, HIDES, and HDS spectrographs.
We examined averaged spectral data of each epoch of both stars and found no
significant time variation in the line peak intensity, central wavelength, and width 
of the  Si~{\sc ii}  emission line at 6239.6 \AA. These data strongly suggest that 
the appearances of the emission line in these two stars are not  temporal phenomena 
but they are  stationary features lasting for at least  a few decades.

Figure 4 clearly demonstrates that very faint WELs can be detected in a relatively
fast rotating star such as $\gamma$ Ori when using high SN and high resolution data.
We can see from this figure that  the emission lines show  single-peaked and 
symmetric profiles in all three stars and that their line 
centers have no shift in wavelength with respect to those of low excitation 
absorption lines of Si~{\sc ii}. Measured half widths of the 
emission lines are the same as those of low excitation absorption lines of Si~{\sc ii} 
in all three stars. These observations imply that the faint WELs are formed nearly at the
same location as the absorption lines and their origins are not circumstellar.
Furthermore, the observed  widths of WELs,  which are the same as those of the rotationally
broadened absorption lines, imply that the emitting gas is covering the whole stellar
surface and is co-rotating with the star.   

The mechanism for populating the highly excited states is still now under 
investigation. 
In a review article, \citet{wahlgren2008}  noted that NLTE can be a significant
mechanism for the production of WELs. \citet{sigut1996} carried out a theoretical 
work on near IR region WELs of Mg~{\sc ii}  in A and B-type stars and published 
predictions of line profiles of four  Mg~{\sc ii} lines between 1.01 $\mu$m and 
4.76  $\mu$m. No observational confirmation
of these Mg~{\sc ii} emission lines has been published yet.
Sigut (2001a, b)  carried out analyses of emission lines of Mn~{\sc ii}  
observed in 3 Cen A and 46 Aql and concluded that these emission lines can be 
 explained by interlocked NLTE  effect combined with the vertical stratification. 
Wahlgren and Hubrig (2000, 2004) proposed an alternative mechanism that the
population of highly excited states might be due to excitation from the
far-UV continuum radiation.

Our present sample covers a different temperature domain 
from that containing the previously known WELs stars on the HR diagram,
and contains only one CP star (HD 133518: B2 IVp He-strong) 
and none of the remaining  sample stars are known  to show spectroscopic anomalies. 
Thus, it seems difficult to postulate a stratification of some specific elements,  
such as Si and Al,  in their atmospheres.  Thus, our results may have a significant bearing
upon our understanding of the outermost atmospheric regions
of B-type stars and the formation of the  abundance anomalies observed
in CP stars, which is generally attributed to the diffusion processes.    
Finally, we notice in table 1 that   
12 stars  belong to the $\beta$ Cep type, the SPB type, or the 
hybrid type pulsational variable stars.  The observed frequency of these variable stars among 
B-type stars is steadily increasing in recent years. Furthermore, weak magnetic fields 
have recently been detected in some of these variable stars \citep{wade2016}.

We plan to carry out spectroscopic observations of early B-type stars in order to 
increase the number of sample stars. One of our interest is observing hot stars
with spectral sub-types between B1 and B2, where we have only a few objects 
in the present sample.  It might be possible to determine the hot end of the region
in the  {\it $T_{\rm eff}$}  range between 23000 K and 27000 K, in which  the 
Si~{\sc ii} line at 6239.6 \AA~ appears as an emission line.  We also plan to observe
 wavelength regions not included in the present study in order to survey for  new
emission features. These observations will provide useful boundary conditions or
constraints to be incorporated in theoretical interpretations.  

\vskip 3mm

This research has made use of the SIMBAD database, operated
by CDS, Strasbourg, France. We thank Dr. E. Kambe for comments and 
Mr. Y. Notsu for his help in preparation of figure materials.




\setcounter {table} {0}
\begin{table*}
      \caption{Data of analysed stars}\label{second}
\tiny
      \begin{center}
      \begin{tabular}{ccccccccccccccc}
\hline\hline
HD	&	Name	&	MK type	& Variable type 	& $\it v$ sin $\it i$	& Instrument	& Obs. Date  & SN & {\it $T_{\rm eff}$} & log $\it g$ &	Ref. &	Si~{\sc ii} 6239 &  & Al~{\sc ii} 6243  &   \\
        &               &               &               &  km s${}^{-1}$   &     & J. D. 2450000 + &    &   K  &  cm s${}^{-2}$           &      & E.W.  &  C.I.   &   E.W.  & C.I.          \\            
(1)	&	(2)	&	(3)	&	(4)	&	(5) 	&	(6)	&   (7)    & (8)  &	(9)	& (10)	& (11)	& (12)	& (13)	& (14)  &  (15)  \\
\hline
205021	&	$\beta$ Cep	&	B1 IV	& $\beta$ Cep$^{a}$     &       20	& 1,3	        &  3257	& 800	&	27000	&	4.05	&	1	& *     & 1.000 	&  *   &	 0.999 \\
36591	&	HR 1861		&	B1 IV	&		        &	9	&	3	&  3322	& 350	&	27000	&	4.12	&	1	& *     & 1.005 	& *    &	 0.997 \\
216916	&	16 Lac		&	B2 IV	& $\beta$  Cep$^{a}$ 	&	10	&	2,3	&  2194	& 520	&	23000	&	3.95	&	1	& --15	& 1.040 	& --3  & 1.008 	\\
214993	&	12 Lac 		&	B2 III	&	Hybrid$^{b}$	&	30	&	3	&  3015	& 440	&	23000	&	3.6	&	3	& --11	&  1.012 & *    & 0.999 	\\
163472	&	V2052 Oph	&	B2 IV-V	& $\beta$  Cep$^{a}$ 	&	75	&	3	&  2902	& 360	&	22490	&	3.95	&	4	& --14	& 1.007 & *    & 0.998 	\\
35468	&	$\gamma$ Ori	&	B2 III	&		        &	52	&	3,4	&  2538	& 410	&	22000	&	3.6	&	3	& --22  & 1.011 & --4  & 1.002	\\
29248	&	$\nu$ Eri	&	B2 III	&	Hybrid$^{b}$	&	32	&	1	&  4030	& 380	&	22000	&	3.85	&	1	& --12	& 1.006 & *    & 1.001 	\\
886	&	$\gamma$ Peg	&	B2 IV	&	Hybrid$^{b}$	&	0	&	1,3	&  4026	& 940	&	22000	&	3.95	&	1	& --19	& 1.057	& --6  & 1.020	\\
16582	&	$\delta$ Cet	&	B2 IV	& $\beta$  Cep$^{a}$ 	&	5	&	3	&  1712	& 430	&	21250	&	3.8	&	1	& --20  & 1.042 & --6  & 1.013	\\
3360	&	$\zeta$ Cas	&	B2 IV	&	SPB$^{c}$ 	&	23	&	1,3	&  3301	& 650	&	20750	&	3.8	&	1	& --21  & 1.027 & --8  &  1.008      \\
35708	&	o Tau		&	B2.5 IV	&		        &	10	&	1	&  4030	& 460	&	20700	&	4.15	&	1	& --15	& 1.020  & --8  & 1.008	\\
35039	&	o Ori		&	B2 IV-V	&		        &	11	&	1	&  4030	& 520	&	19600	&	3.56	&	1	& --20	& 1.048  & --9  & 1.020	\\
42690	&	HR 2205		&	B2 V	&		        &	0	&	1	&  4030	& 410	&	19299	&	3.81	&	2	& --17	& 1.030  & --7  & 1.011	\\
170580	&	HR 6941		&	B2V	&	Hybrid$^{b}$	&	0	&	2	&  3184	& 970	&	19175	&	4.02	&	5	& --13	& 1.039  & *    & 1.001 	\\
133518	&	HIP 73966	&	B2 IVpHe &                      &	0	&	4	&  1983 & 270	&	19000	&	4.0	&	6	& --10 	& 1.053  & --5  & 1.024	\\
32249	&	$\psi$ Eri	&	B3 V	&		        &	30	&	1	&  4027	& 510	&	18890	&	4.13	&	2	& --4:	& 1.005 & *    & 1.000 	\\
34447	&	HR 1731		&	B3 IV	&		        &	10	&	1	&  4027	& 270	&	18480	&	4.10	&	2	& --8	& 1.021 & *    & 1.000  	\\
160762	&	$\iota$ Her	&	B3 IV	&	SPB$^{d}$ 	&	0	&	1,2,3	&  4026	& 610	&	17500	&	3.8	&	1	& --9	& 1.030  & +3   & 0.990 	\\
196035	&	HR 7862		&	B3 IV	&		        &	20	&	1	&  4028	& 360	&	17499	&	4.36	&	2	& +3:	& 0.998 & +3:  & 0.996 	\\
43157	&	HR 2224		&	B5 V	&		        &	30	&	1	&  4027	& 410	&	17486	&	4.12	&	2	& +9:	& 0.983 & +5:  & 0.993 	\\
223229	&	HR 9011		&	B3 IV	&		        &	30	&	1	&  4026	& 370	&	17327	&	4.20	&	2	& --8:	& 1.005 & *    & 0.999        \\
89587	&	HIP 50519	&	B3 III	&		        &	no data	&	4	&  2217	& 540	&	17000	&  no data		&	7	& --6	& 1.016  & +3   & 0.988 	\\
176502	&	V543 Lyr	&	B3 V	&		        &	0	&	1	&  4027	& 390	&	16821	&	3.89	&	2	& +6	& 0.986 & +9   & 0.972       \\
41753	&	$\nu$ Ori	&	B3 V	&		        &	30	&	1	&  4029	& 410	&	16761	&	3.9	&	2	& --4	& 1.005 & *    & 0.994 	\\
25558	&	40 Tau		&	B3 V	&  SPB$^{d}$	        &	30	&	1	&  4028	& 390	&	16707	&	4.29	&	2	& +9	& 0.982 & +5:  & 0.992 	\\
44700	&	HR 2292		&	B3 V	&		        &	0	&	1	&  4029	& 340	&	16551	&	4.29	&	2	& +6	& 0.982 & +6   & 0.972 	\\
186660	&	HR 7516		&	B3 III	&		        &	0	&	1	&  4027	& 310	&	16494	&	3.57	&	2	& --6	& 1.025 & +6   & 0.976 	\\
181858	&	HR 7347		&	B3 II-III&  	                &	0	&	1	&  4027	& 270	&	16384	&	4.19	&	2	& +9:	& 0.979 & +8:  & 0.984 	\\
184171	&	8 Cyg		&	B3 IV	&		        &	15	&	1	&  4027	& 430	&	15858	&	3.54	&	2	& +6	& 0.988 & +6   & 0.989 	\\
198820	&	HR 7996		&	B3 III	&		        &	15	&	1	&  4027	& 280	&	15852	&	3.86	&	2	& +6	& 0.989 & +7:  & 0.989 	\\
209008	&	18 Peg		&	B3 III	&	SPB$^{e}$        &	5	&	1	&  4028	& 360	&	15800	&	3.75	&	1	& +9	& 0.980 & +12  & 0.982 	\\
28375	&	 HR 1415	&	B3 V	&		        &   	0	&	1	&  4027	& 380	&	15278	&	4.30	&	2	& +30	& 0.950 & +20  & 0.971 	\\
11415	&	$\epsilon$ Cas	&	B3 III	&		        &	30	&	1,3	&  4027	& 650	&	15174	&	3.54	&	2	& +10	& 0.984 & +14  & 0.986 	\\
147394	&	$\tau$ Her	&	B5 IV	&		        &	30	&	1	&  4026	& 610	&	14898	&	4.01	&	2	& +27	& 0.970 & +18  & 0.980        \\
17081	&	$\pi$ Cet	&	B7 V	&		        &	25	&	1	&  4029	& 710	&	13063	&	3.72	&	2	& +37	& 0.944 & +20  & 0.966 	\\
\hline
        \end{tabular}
     

(1) HD number.  (2) Star name. (3) Spectral type. (4) References:  
a:  \citet{stankov2005}, b: \citet{moravveji2016}, c: \citet{briquet2016}, 
d: \citet{szewczuk2015}, and e:  \citet{Irrgang2016}. 
 (5) Rotational velocity ($\it v$ sin $\it i$).  (6) Used spectrographs. 1: HIDES, 2: HDS, 3: Elodie, 
4: UVES. 
(7) Date of observation in J. D. When multiple observations are available, date of observations
used in measurements (12) -- (15) are given.
(8) SN ratio measured at the continuum level near 6250 \AA. (9) Effective temperature . (10) Logarithm 
of surface gravity.  (11)  Sources of {\it $T_{\rm eff}$} and log $\it g$. 1: \citet{nieva2013}, 2: \citet{takeda2010},
 3: \citet{lefever2010}, 4: \citet{aerts2014}, 5: \citet{morel2008}, 6: \citet{alecian2014},
7: Estimated using the S~{\sc ii}  and N~{\sc ii}  line intensity ratio. (12) and (14) Equivalent widths (E. W.) of the 
Si~{\sc ii}  6239.6 and the Al~{\sc ii}  6243.4 in m\AA. Data of emission and absorption features are given in negative and positive numbers, respectively. Entries marked with a colon(:) are less accurate. An asterisk (*)  indicates that no emission or absorption feature can be recognized.
(13) and (15) Central intensities (C. I.) of the Si~{\sc ii}  6239.6 and the Al~{\sc ii}  6243.4 lines
relative to the local continuum level.


 \end{center}
\end{table*}

\setcounter {table} {1}\begin{table}
      \caption{Emission lines of Al~{\sc ii}  and Si~{\sc ii}  observed in $\gamma$ Peg}\label{first}
\tiny
      \begin{center}
      \begin{tabular}{cccccccc}
\hline\hline
 Ion  & Mult. No. & $\lambda$  & Excitation potential & Configuration &  & log $\it gf$  & Peak intensity  \\
        &   &   (\AA)    & (eV)          & Upper   &  Lower &     &                 \\
\hline
Al~{\sc ii}   &   --  & 5593.30  & 13.26    & 	 3$\it s$4$\it p$ &  3$\it s$4$\it d$  & 0.337	 & 1.003      \\
              &	10    & 6226.18  & 13.07   & 	3$\it s$4$\it p$ & 3$\it s$4$\it d$  & 0.037  & 1.004      \\
              &	10    & 6231.75  & 13.07    &  3$\it s$4$\it p$ 	& 3$\it s$4$\it d$   &   0.389	 & 1.014      \\
              &  10   & 6243.36  & 13.08    &  3$\it s$4$\it p$ 	&  3$\it s$4$\it d$  &   0.659    & 1.020      \\
Si~{\sc ii}   & --    & 5185.52  & 12.84    & 3$\it s$$^{2}$4$\it f$     &  3$\it s$$^{2}$7$\it g$   & -0.302     & 1.013      \\
              & --    & 5185.56  & 12.84    & 3$\it s$$^{2}$4$\it f$     &  3$\it s$$^{2}$7$\it g$      & -0.456     & 1.013      \\
              & --    & 5466.43  & 12.53    & 3$\it s$$^{2}$4$\it d$    & 3$\it s$$^{2}$6$\it f$       & -0.237     & 1.016      \\  
              &--     & 5466.89  & 12.53  &  3$\it s$$^{2}$4$\it d$     & 3$\it s$$^{2}$6$\it f$       & -0.082     & 1.012      \\
              &--     & 5688.81  & 14.19  & 3$\it s$3$\it p$($^{3}$P$^{\circ}$)3$\it d$ & 3$\it s$3$\it p$($^{3}$P$^{\circ}$)4$\it p$ & 0.126   & 1.009      \\ 
              &--     & 5701.37  & 14.17  & 3$\it s$3$\it p$($^{3}$P$^{\circ}$)3$\it d$ & 3$\it s$3$\it p$($^{3}$P$^{\circ}$)4$\it p$ & -0.057  & 1.004      \\
              &--     & 6239.61  & 12.84  & 3$\it s$$^{2}$4$\it f$   & 3$\it s$$^{2}$6$\it g$     & 0.177	& 1.057      \\
              &--     & 6239.66  & 12.84  & 3$\it s$$^{2}$4$\it f$    &  3$\it s$$^{2}$6$\it g$            & 0.021      & 1.057      \\
\hline
        \end{tabular}

Multiplet numbers are taken from \citet{moore1959} and
atomic data (wavelengths $\lambda$,  excitation potentials, electron configurations and  log $\it gf$ 
values) are taken from the NIST atomic database. 
 
      \end{center}
\end{table}


\begin{figure}
     \begin{center}
       \FigureFile(70mm,80mm){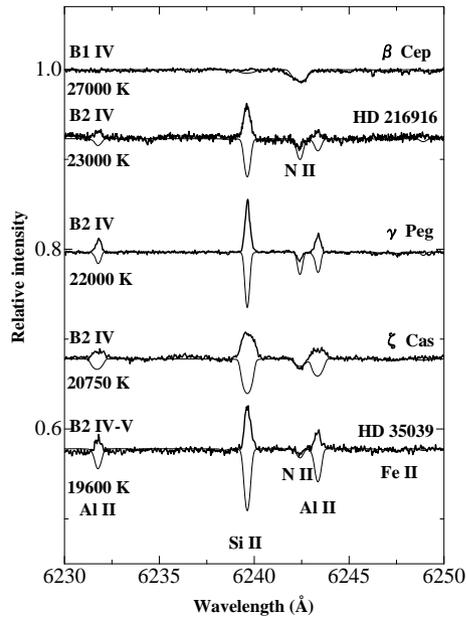}
     \end{center}
     \caption{Sample spectra of five hot (from B1 to B2) B-type stars
                between 6230 and 6250 \AA. Lines of Si~{\sc ii}   at 6239.6 \AA~  
                and of  Al~{\sc ii} at 6231.75 \AA~ and at 6243.37 \AA~ are shown.
                Spectral types and effective temperatures are indicated. 
                Thick and thin lines show observed and simulated spectra,
                respectively. }  
    \end{figure}
\begin{figure}
     \begin{center}
  \FigureFile(70mm,80mm){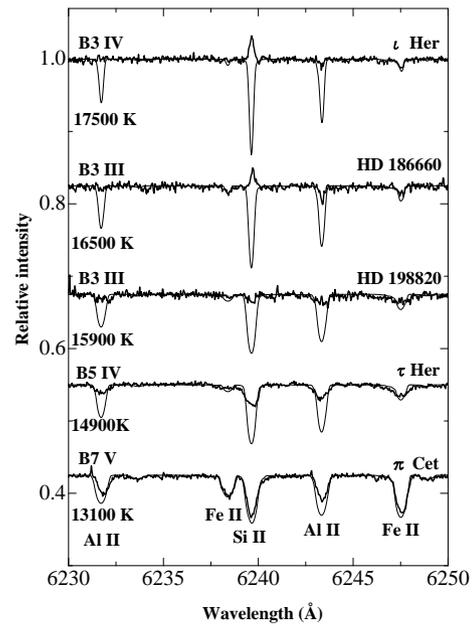}
     \end{center}
     \caption{The same as figure 1, but for five relatively cool  (from B3 to B7) 
                 B-type stars. }
        \end{figure}

\begin{figure}
    \begin{center}
  \FigureFile(65mm,70mm){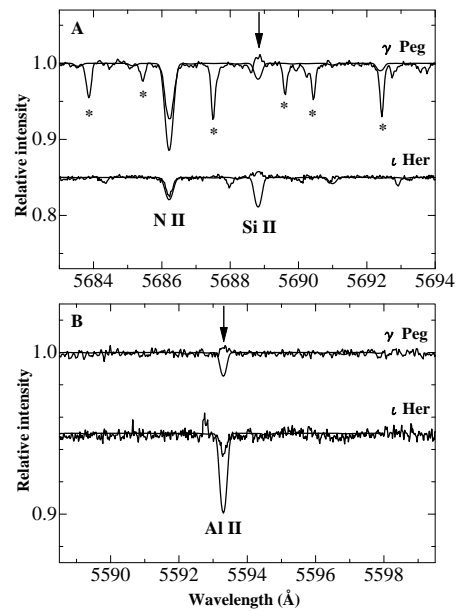}
    \end{center}
     \caption{Si~{\sc ii} line at 5688.8 \AA~ (panel A) and Al ~{\sc ii} 
                line at 5593.3 \AA~ (panel B)  
                observed in $\gamma$ Peg and $\iota$ Her.
                Thick and thin lines show observed and simulated spectra, respectively.
                Asterisks indicate atmospheric absorption lines. }
        \end{figure}
\begin{figure}
    \begin{center}
  \FigureFile(70mm,85mm){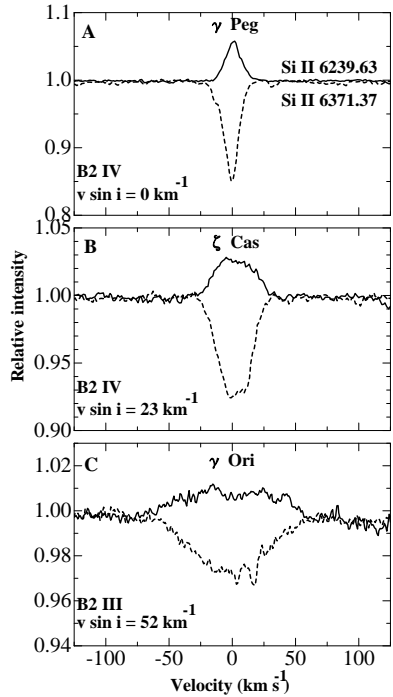}
    \end{center}
     \caption{Profiles of the  Si~{\sc ii} emission line  at 6239.6 \AA~ (thick lines) and 
                those of low excitation Si~{\sc ii} absorption  line  at 6371.37\AA~ (broken lines) 
                for three B2 type stars $\gamma$ Peg (panel A),   $\zeta$ Cas (panel B), 
                and $\gamma$ Ori (panel C).  Spectral types and data of $\it v$ sin $\it i$ are
                indicated.  Profiles are plotted on the velocity scale.  }
        \end{figure}

\end {document}